\documentclass[sort&compress]{elsarticle}

\usepackage{lineno}

\usepackage{graphicx}
\usepackage{dcolumn}
\usepackage{bm}
\usepackage{upgreek}
\usepackage{hyperref}
\usepackage{ams math}
\usepackage{multirow}
\usepackage[margin=1.3in]{geometry}
\let\oldequation\equation
\let\oldendequation\endequation

\renewenvironment{equation}
  {\linenomathNonumbers\oldequation}
  {\oldendequation\endlinenomath}

\usepackage{mdframed} 

\usepackage{multicol} 

\usepackage{nomencl} 

\makenomenclature

\usepackage{hyperref}
\usepackage{soul,color,xcolor}
\sethlcolor{white}
\soulregister{\cite}7
\soulregister{\citep}7
\soulregister{\citet}7
\soulregister{\ref}7
\soulregister{\pageref}7
\usepackage{etoolbox}
\renewcommand\nomgroup[1]{%
  \item[\bfseries
  \ifstrequal{#1}{G}{Greek symbols}{%
  \ifstrequal{#1}{S}{Subscripts}{}}%
]}

\newcommand{\Oh}{O\!h}%
\newcommand{\Bo}{B\!o}%
\renewcommand{\Bo}{\text{Bo}}%
\renewcommand{\Oh}{\text{Oh}}%
\journal{Journal of Colloid and Interface Science}

\begin{document}

\begin{frontmatter}

\title{Bridge evolution during the coalescence of immiscible droplets}

\author{Huadan Xu}
\author{Tianyou Wang}
\author{Zhizhao Che\corref{cor1}}
\cortext[cor1]{Corresponding author.
}
\ead{chezhizhao@tju.edu.cn}
\address{State Key Laboratory of Engines, Tianjin University, Tianjin, 300350, China.}

\begin{abstract}
\emph{Hypothesis}: Droplet coalescence is a common phenomenon and plays an important role in many applications. When two liquid droplets are brought into contact, a liquid bridge forms and expands quickly. Different from miscible droplets, an extra immiscible interface exists throughout the coalescence of immiscible droplets and is expected to affect the evolution of the liquid bridge, which has not been investigated. We hypothesized that the liquid bridge of immiscible droplets exhibits a different growth dynamics.

\noindent \emph{Experiments}: We experimentally study the coalescence dynamics of immiscible droplets. The evolution of the liquid bridge is measured and compared with miscible droplets. We also propose a theoretical model to analyze the effects of immiscibility.

\noindent \emph{Findings}: We find that immiscibility plays different roles in the viscous-dominated and inertia-dominated regimes. In the initial viscous-dominated regime, the coalescence of immiscible droplets follows the linear evolution of the bridge radius as that of miscible droplets. However, in the later inertia-dominated regime, the coalescence of immiscible droplets is slower than that of miscible droplets due to the water-oil interface. By developing a theoretical model based on the force balance, we show that this slower motion is due to the immiscible interface and the extra interfacial tension. In addition, a modified Ohnesorge number is proposed to characterize the transition from the viscous-dominated regime to the inertia-dominated regime.
\end{abstract}

\begin{keyword}
\texttt {
Droplet coalescence \sep
Immiscibility \sep
Liquid bridge \sep
Bridge evolution
}
\end{keyword}

\end{frontmatter}


\def \scaleSize {0.8}
\def \scaleSiz2 {0.6}
\section{Introduction}

Droplet coalescence is a ubiquitous phenomenon both in natural and industrial processes, such as raindrop formation \cite{Whelpdale1971RainDropGrowth}, water-oil separation \cite{Shen2015MicroDropletCoalescence}, spray \cite{Wu2021MixingSpray}, coating \cite{Pathak2021dropletCondensation}, and fuel atomization in combustion engines \cite{Wang2017interdropletSpace}. It has been identified to play an important role in the relevant applications. For example, water removal and oil desalting processes greatly affect petroleum transportation \cite{liu2021DropletParticles}; a dominated emulsion process decides the shelf life of the emulsion-based products \cite{Shimizu2015Immisciblemixture}; and the coalescence speed is crucial to the growth of liquid crystals and the quality of the final product \cite{Klopp2020SelfSimilarity}. In recent years, with the development of microfluidic technologies, the coalescence of droplets in microchips has been designed and controlled to achieve numerous novel functions, such as cellular and subcellular structure analysis \cite{Sohrabi2020dropletApplication}, protein crystallization \cite{Ferreira2018ProteinCrystallization}, as well as the synthesis of nanoparticles \cite{Zhou2016DropletCoalescenceNanoparticle}.

The coalescence of two droplets involves a complex process. When two droplets approach each other, an intervening film of the surrounding phase first forms between the droplets, even if the surrounding phase is a gas. When the intervening film is thin enough, the van der Waals force and other intermolecular forces begin to dominate \cite{Boyson2007OilCoalescence}, then a thin \hl{film of surrounding fluid } ruptures, and the coalescence of the two droplets is initiated. After that, a liquid bridge forms between the two droplets and expands laterally \cite{Paulson2011DropletCoalescence}. The large curvature formed at the bridge results in strong Laplace pressure driving the droplet fluid into the bridge region, leading to the expansion of the bridge. As the bridge expands over time, two regimes have been identified depending on whether the viscous force or the inertia is the dominant force resisting the surface tension force, i.e., a viscous-dominated regime and an inertia-dominated regime \cite{Xia2019dropletCoalescence}. The two regimes can be described by different power laws. For miscible droplets, the growth of the bridge radius in the viscous-dominated regime follows $r^* \sim t^*$ \cite{Paulson2011DropletCoalescence}, where $r^*$ is the nondimensional bridge radius r scaled by the initial droplet radius $R_0$, and $t^*$ is the nondimensional time scaled by the viscocapillary time scale  ${{t}_{\nu }} \equiv {\mu {{R}_{0}}}/{\sigma }$, with $\mu$ and $\sigma$ being the viscosity and the surface tension of the droplet. In the later inertia-dominated regime, the bridge evolution is often described by the 1/2 power-law scaling, $r^* \sim  t^{*1/2}$, where $t^*$ is the nondimensional time scaled by the capillary-inertial time scale ${{t}_{\sigma }} \equiv \sqrt{{\rho R_{0}^{3}}/{\sigma }}$ \cite{Wu2004scalinglawCoalescence} with $\rho$ being the density of the droplet fluid. Apart from the viscous-dominated and the inertia-dominated regimes, an inertially limited viscous regime was proposed by Eggers et al.\ \cite{Eggers1999DropletCoalescence}, which is suggested to follow the form $r^* \sim t^* \ln(t^*)$, with time scaled by the viscocapillary time scale ${{t}_{\nu }}$. However, many experiments only show a linear relationship \cite{Aarts2005DropletcoalescenceHydrodynamics, Burton2007Dropletpinchoff, Paulson2011DropletCoalescence, Thoroddsen2005CoalescenceSpeed} and the logarithmic correction $\ln(t^*)$ was not observed.

The coalescence of droplet pairs of identical fluid has been widely investigated, including two highly viscous droplets \cite{Rahman2019ViscousCoalescence}, inviscid droplets \cite{Duchemin2003InviscidCoalescence}, equal-sized droplets \cite{Sprittles2014InertialCoalesence}, and un-equal sized droplets \cite{Deka2019dropletcoalescence}. Their coalescence processes are governed by the viscous forces, capillarity, and inertia, and are in accordance with the above-mentioned power laws. Besides, due to the rising applications of droplet coalescence such as droplet-based biochemical reactions \cite{Shen2015MicroDropletCoalescence, Sohrabi2020dropletApplication}, coalescence of droplet pairs of different fluids is gaining increasing attention, which involves the coexistence of multiple components, thus making the coalescence process more complex than the single-component droplet coalescence \cite{Nowak2016surfactantDropCoalescence, Shen2015MicroDropletCoalescence}. As the surface tension is the driving force in the coalescence process, several studies focused on the influence of the surface-tension difference between the two coalescing droplets. Thoroddsen et al.\ \cite{Thoroddsen2007InitialDropletCoalescence} experimentally investigated the coalescence of two droplets with different surface tension, and proposed that the liquid with lower surface tension controls the coalescence speed. In addition, adding surfactants could also lead to the spatial variation of surface tension and affect the interface deformation. Martin and Blanchette \cite{martin2015DropsAndBubbleCoalescence} numerically investigated the effects of surfactants on coalescence dynamics and found that surfactants can reduce the critical Ohnesorge number for partial coalescence. In the experimental work by Nowak et al.\ \cite{Nowak2016surfactantDropCoalescence}, they found that the diameter of the liquid bridge decreased as the surfactant concentration increased. Recently, Amores et al.\ \cite{Constante2021MarangoniStressdropletcoalescence} numerically studied the effect of Marangoni stress on the coalescence of surfactant-laden droplets, and explained the cause of the pinch-off inhibition \cite{martin2015DropsAndBubbleCoalescence} from the perspective of vortex advection. Apart from Newtonian fluids, \hl{Varma et al. \cite{Varma2022Rheocoalescence} considered the coalescence of polymeric droplets and identified three regimes depending on polymer concentrations, and then established a universal model describing the expansion of the bridge radius \cite{Varma2020PolymericFluidsCoalescence}.}

In the above studies of the droplet coalescence of multi-component fluids, the investigated fluid pairs are all miscible with each other. The coalescence process will become more complex when the two droplets are immiscible. Nonetheless, the studies concerning immiscible droplets are limited. For the reported studies involving the coalescence of immiscible droplets, some focused on the film drainage during the coalescence of a droplet with an immiscible liquid pool \cite{Bozzano2010InterfaceCoalescence}, while others reported the different morphologies of compound droplets at their equilibrium state \cite{Bansal2017SessileDropletoscillation}. \hl{Another related study by Bansal et al.\ \cite{Bansal2018electroCoalescence} investigated the coalescence of compound droplets induced by electrowetting, and observed two regimes of coalescence and non-coalescence.} To better understand and control the droplet coalescence and facilitate the relevant applications, we experimentally investigate the coalescence of two immiscible droplets. The bridge evolution dynamics is studied and compared with that of miscible droplets. The influence of the immiscibility between the droplets on the evolution of the liquid bridge is analyzed. The viscous-dominated regime, the inertia-dominated regime, and their transition is investigated.

\section{Experimental method}\label{sec:2}
\subsection{Droplet properties}\label{sec:21}
The liquids used in the experiments can be divided into two groups. For immiscible droplet pairs, we used water and silicone oil as the water and oil phases, respectively. Besides directly using pure silicone oils of different viscosities (3 -- 50 cSt), we also prepared oil droplets with intermediate viscosities by mixing silicone oil of 5 cSt and 500 cSt at different ratios. Their viscosities are measured with a viscometer (NDJ-5S-8SPro) at 20 $^\circ$C with deviations around $ \pm 10 \%$ compared to the measurements obtained by Zhang et al.\ \cite{Zhang2018PolydimethylsiloxaneMixturesWetting}, and the densities were obtained by weighing a known volume of mixture liquids. For miscible droplet pairs, to obtain a similar viscosity and surface tension as silicone oil, we used ethanol-glycerol mixtures of different concentrations as the oil phase. The properties of the droplet fluids are listed in Table \ref{tab:1}. The viscosity ratio $\mu^*$ is defined as the ratio of the viscosity between the silicone oil and water for the immiscible droplet pairs $\mu ^* \equiv {{{\mu }_{\text{silicone-oil}}}}/{{{\mu }_{\text{water}}}}$, or between the ethanol-glycerol mixture and water for the miscible droplet pairs $\mu ^* \equiv {{{\mu }_{\text{ethanol-glycerol}}}}/{{{\mu }_{\text{water}}}}$. The experiments were carried out at the temperature of 20 $^\circ$C and atmospheric pressure with an air density of $\rho_g = 1.205$ kg/m$^3$ and an air viscosity of $\mu_g = 0.0181$ mPa$\cdot$s. The room humidity was kept over 30\% to avoid the electrostatic effect.

\begin{table}
\caption{Properties of the liquids used in this study. The values in the parentheses for the silicone oils are the kinematic viscosity. The properties of the ethanol-glycerol mixtures are from Ref. \cite{Alkindi2008EthanolGlycerolMixture}, \hl{the properties of the silicone oils are from Ref. \cite{Aryafar2006InertiaCoalescence} and} the surface tension of silicone oil mixtures is from Ref. \cite{Zhang2018PolydimethylsiloxaneMixturesWetting}. SOM is the silicone oil mixture (by mixing silicone oil of 5 cSt and 500 cSt) and the values in the parenthesis are the weight concentration of the 500 cSt silicone oil. EGM is the ethanol-glycerol mixture and the values in the parenthesis are the weight concentration of glycerol. \hl{The interfacial tension between oil and water is 35 mN/m according to previous measurements \cite{Brassard2008ElectroMicrofluidic}}.}\label{tab:1}
\centering
\begin{tabular}{p{0.2\textwidth}p{0.15\textwidth}p{0.15\textwidth}p{0.18\textwidth}}
\hline
Liquid  & Viscosity & Density   & Surface tension\\
& $\mu$ (mPa$\cdot$s)  &$\rho$ (kg/m$^3$)  &$\sigma$ (mN/m)\\ \hline
Water                   & 1.0       & 996             & 72.0   \\
Silicone oil (3 cSt)    & 2.67      & 890             & 18.9   \\
Silicone oil (5 cSt)	& 4.56		& 913		& 18.9  \\
Silicone oil (10 cSt)	& 9.35		& 935		& 19.4  \\
Silicone oil (20 cSt)	& 18.9		& 949		& 19.8  \\
Silicone oil (50 cSt)	& 48.0		& 960		& 20.4  \\
SOM (8 wt\%)			& 8.768		& 917.6		& 18.2  \\
SOM (10 wt\%)			& 12.69		& 918.7		& 18.2  \\
SOM (15 wt\%)			& 16.23		& 921.6		& 18.2  \\
SOM (25 wt\%)			& 25.32		& 927.3		& 18.2  \\
EGM (33.55 wt\%)		& 5.22		& 916.05	& 23.5  \\
EGM (40.62 wt\%)		& 9.09		& 946.41	& 23.5  \\
EGM (56.67 wt\%)		& 20.31		& 1020.27	& 25.1  \\
EGM (68.06 wt\%)		& 54.01		& 1077.53	& 28.5	\\
\hline
\end{tabular}
\end{table}

\subsection{Coalescence configurations}\label{sec:22}
\begin{figure}
  \centering
  \includegraphics[scale=0.75]{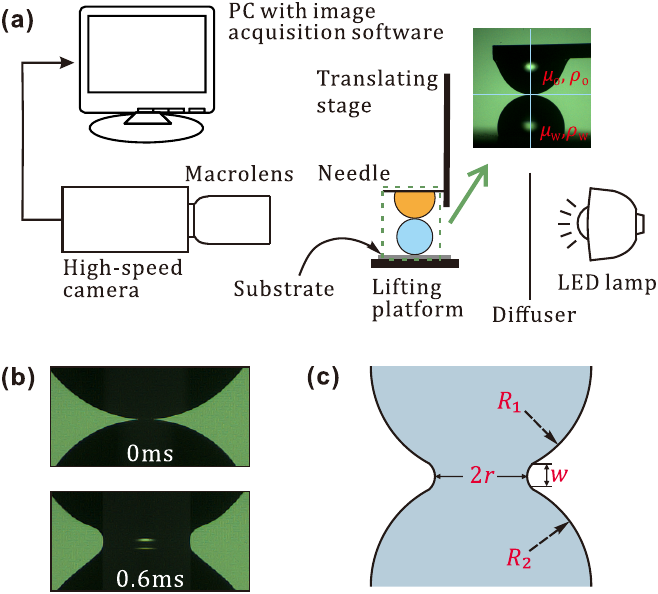}
  \caption{(a) Experimental setup for droplet coalescence. (b) Typical images before and after the coalescence. (c) Schematic illustration of the droplet coalescence. $R_1$ and $R_2$ are the \hl{local radii of curvature} of the two droplets, $r$ and $w$ denote the radius and width of the liquid bridge, respectively.
}\label{fig:1}
\end{figure}

The experimental system is schematically shown in Figure \ref{fig:1}a. To probe microscopic details of the coalescence dynamics, \hl{which is the bridge evolution in the present study}, we used vertically aligned droplet pairs, similar to the droplet configuration used by Varma et al.\ \cite{Varma2020PolymericFluidsCoalescence}. At the bottom is a sessile droplet of water on a hydrophobic substrate, which was obtained by treating a silica glass with Fluorosilicone polymer to achieve an equilibrium contact angle of about $100^\circ$--$120^\circ$. At the top is a pendent droplet of the lower-surface-tension fluid (silicone oil for immiscible droplet pairs and ethanol-glycerol mixture for miscible droplet pairs) and was pre-positioned on a thin needle. \hl{Two different needles were used, and their outer diameters are 0.23 and 0.24 mm, respectively.} The two droplets were then brought into contact by slowly ascending the sessile droplet through a three-axis translation stage while fixing the position of the pendent droplet. In this way, we could better control the droplet size and avoid the influence of the additional momentum when introducing liquid into the droplet. \hl{It should be noted that, the needle diameter is much smaller than that of the droplet, and the surface wettability of the needle was modified to change its surface energy to prevent the oil droplet from spreading too far on the needle wall. The present study concerns the initial stage of droplet coalescence, which is often considered as `local' and is independent of the geometry far from the contact point. All these factors make sure that the droplet configuration in this experiment is suitable to investigate the short-time dynamics of liquid bridge growth.} In addition, the needle and the substrate were grounded electrically to avoid the interference of naturally accumulated charges on the droplet surface \cite{Yokota2011CoalescenceCrossover}.

The influence of gravity on the coalescence can be estimated by the Bond number, $\Bo \equiv {\rho {{R}^{2}}g}/{\sigma }$, which represents the ratio between the gravitational force and the surface tension force. The Bond numbers for the higher-surface-tension droplet and the lower-surface-tension droplet are about 0.15 and 0.2, respectively. Therefore, the effect of gravity due to the vertical droplet arrangement can be neglected in the coalescence process.

\subsection{Visualization and image processing}\label{sec:23}
The rapid expansion of the liquid bridge was captured by a high-speed camera (Phantom V1612) at 250,000 fps (frames per second). Close-up views were accomplished with a microscopic lens (Navitar Zoom 6000) with illumination using a light-emitting diode (LED) light source. This setting allowed for temporal and spatial resolutions of about 4 $\mu$s and 6 $\mu$m, respectively. To avoid the droplet evaporation due to the heating from the illumination, we added a short pass filter on the diffusion glass to eliminate the far-infrared light (wavelength longer than 40 $\mu$m) of the LED light source. \hl{Typical images of a sessile water droplet and a pendent oil  droplet before and after the coalescence are shown in Figure 1b}. The evolution of the bridge radius was tracked from the experimental images using an edge-detection algorithm, which adopted a threshold for segmentation according to the distinction in the brightness between the droplet and the background. In some experiments, to enhance the contrast between the two droplets, the water droplet was dyed red by adding Rhodamine B with a concentration lower than 0.1 wt\%, which was tested to have not induced significant changes in the physical properties of the droplet \cite{Lu2020DropletImpactFilm}.

\section{Results and discussion}\label{sec:3}
\subsection{Coalescence process}\label{sec:31}
\begin{figure}
  \centering
  \includegraphics[scale=0.8]{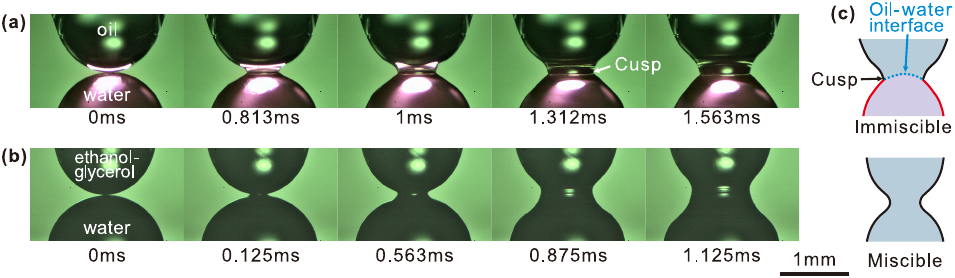}
  \caption{Image sequences of droplet coalescence. (a) Image sequence for the coalescence of two immiscible droplets (a water droplet and a silicone oil droplet). The viscosity ratio $\mu^*$ is 9.35. The water droplet was dyed red by adding Rhodamine B to enhance the contrast of the water-oil interface. (b) Image sequence for coalescence of two miscible droplets (a water droplet and a droplet of ethanol-glycerol mixture). The viscosity ratio $\mu^*$ is 9.09. The video clips for the two processes are available as Supplemental Material. (c) Illustration of the bridge after the coalescence for immiscible and miscible droplet pairs.}\label{fig:2}
\end{figure}

The snapshots for the coalescence of immiscible and miscible droplets are compared in Figure \ref{fig:2}. After two droplets contact each other, they, under the van der Waals attraction \cite{Sulaimon2018WaterOilEmulsion}, contact quickly by forming a thin liquid bridge. We can observe that the bridge region exhibits a cusp where the surface of the two immiscible droplets meets, as shown in Figure \ref{fig:2}a, and this bridge shape is different from the miscible coalescing droplets, in which there is no such cusp, as illustrated in Figure \ref{fig:2}c. This difference in the surface shape can be attributed to the miscibility difference. For the miscible droplet pair, instantaneous interfacial tension would vanish during the coalescing process, which leads to a smooth liquid bridge region \cite{Thoroddsen2007InitialDropletCoalescence}. For the immiscible droplet pair, the water-oil interface and the water-oil interfacial tension always exist between the two coalescing droplets, thus causing a cusp on the liquid bridge surface, as shown in Figure \ref{fig:2}a. Similar cusps also exist at the wetting ridge when a water droplet is deposited on an oil-infused surface \cite{Smith2013dropletmobility}. In addition, the water-oil interfacial tension can act as a recoiling force during the impact of a water droplet onto an oil film, and shorten the crown ascending stage \cite{Bernard2020DropletImpactFilm}. Therefore, the evolution of the water-oil interface and the interfacial tension force plays an important role during the expansion of the liquid bridge of two immiscible droplets.

\begin{figure}
  \centering
  \includegraphics[scale=0.85]{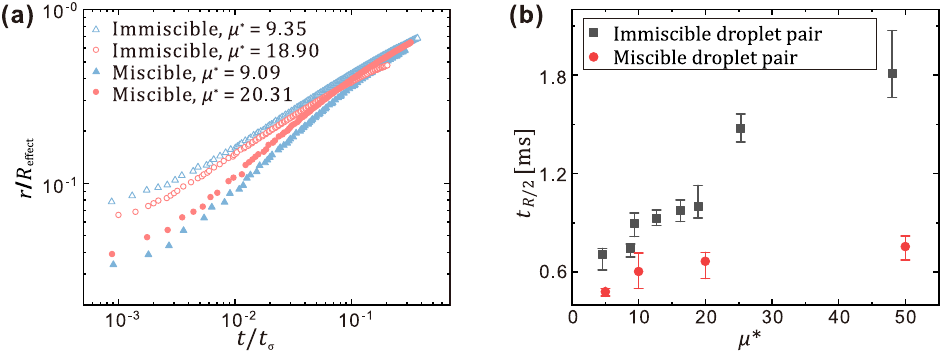}
  \caption{(a) Evolution of liquid bridge radius after the coalescence for miscible and immiscible droplet pairs. (b) Bridge expansion time ${{t}_{{R}/{2}}}$ that the bridge radius grows to ${{{R}_{\text{effect}}}}/{2}$ under different viscosity ratios. \hl{The error bars are the deviations from the average value of five repeated tests using the same liquids.}}\label{fig:3}
\end{figure}

To quantitatively compare the bridge evolution for miscible and immiscible droplet pairs, we first measure the evolution of the bridge radius from the high-speed images by image processing. The bridge radius is defined as the minimum radial distance of the liquid bridge from the surface to the axis, as illustrated in Figure \ref{fig:1}c. The bridge radius is plotted against the dimensionless time for miscible and immiscible droplet pairs in a log-log scale, see Figure \ref{fig:3}a. The time is scaled by the capillary-inertia timescale $t_\sigma$. For the coalescence of miscible droplet pairs, the characteristic timescale $t_\sigma$ can be expressed as ${{t}_{\sigma }} \equiv \sqrt{{{{\rho }_{\text{avg}}}{{R}_{\text{effect}}^{3}}}/{\sigma }}$, where ${{\rho }_{\text{avg}}} \equiv \left( {{\rho }_{\text{w}}}+{{\rho }_{\text{o}}} \right)/2$ is the average density of the two droplets, and ${{R}_{\text{effect}}} \equiv {2}/{\left( {1}/{{{R}_{1}}} +{1}/{{{R}_{2}}} \right)}$ is the effective radius of the droplet pairs \cite{Thoroddsen2007InitialDropletCoalescence}. \hl{${{R}_{1}}$ and ${{R}_{2}}$ are defined as the local radii of the two droplets considering the slight shape deviation from a perfectly spherical shape, and were obtained by fitting the edge of the droplet near the contact point via image processing using ImageJ  \cite{Schindelin2012IMangeJ}.} Since the coalescence speed is determined by the lower-surface-tension droplet for miscible droplets \cite{Thoroddsen2007InitialDropletCoalescence}, the lower surface tension of the droplet pair is used in the above miscible timescale. For the immiscible droplet pairs, the characteristic timescale $t_\sigma$ can be expressed as ${{t}_{\sigma }}=\sqrt{{{{\rho }_{\text{avg}}}R_{\text{effect}}^{3}}/{{{S}_{\sigma }}}}$, where ${{S}_{\sigma }}={{\sigma }_{wa}}-({{\sigma }_{wo}}+{{\sigma }_{oa}})$ is a spreading parameter which can be used to describe the spreading in three-phase systems \cite{Gennes2004Capillaritywetting}.

It can be noted that the bridge radius in Figure \ref{fig:3}a does not start from zero. This is not only due to the limitation of digital image analysis in the experimental measurement, but more importantly due to the intervening air layer between the two droplets in the coalescence process. When the two droplets approach each other, they do not contact immediately but form an intervening air layer between the two droplets. The air layer can be at a submicrometer scale in thickness, which inhibits the further drainage of the air layer. The lubricating flow in the thin air layer increases the local pressure in the air layer and deforms the shape of the surfaces. Therefore, at the moment of the direct contact of the two droplets, the droplet surface has been deformed, which affects the initial radius of the liquid bridge. It can be found that the initial bridge radius of the immiscible droplets is larger than that of the miscible droplets, which can be attributed to the delayed drainage of the air layer due to the surface-tension difference between the two droplets, as reported by Choi and Lee \cite{Choi2014FilmDrainage} in their numerical work. Moreover, for the immiscible droplet pair, the van der Waals force is relatively weak between water-oil surfaces \cite{Boyson2007OilCoalescence}, which could be another reason for the larger initial deformation. Since there have been many studies on the drainage of the intervening air layer between two liquids \cite{Bozzano2010InterfaceCoalescence, Chan2011DropsBubbleCoalescence, Choi2014FilmDrainage}, we do not study the details of the drainage, but rather focus on the bridge evolution after the collapse of the air layer. \hl{In our experiments, the initial radius of the liquid bridge is limited to be less than $8\%$ of the droplet radius. For some cases with an unusually large initial radius (e.g., larger than $10\%$ of the droplet radius, which may be due to the asymmetrical contact), they were abandoned to minimize the uncertainty. Therefore, we can rule out the influence of the initial radius in the present study. In addition, to make sure we could capture the whole event, we started recording long before the contact of the two droplets.} We then measure the bridge evolution by starting from the moment when we could first detect a discernible increase in the bridge radius, which is a common practice in many studies of droplet coalescence processes \cite{Rahman2019ViscousCoalescence, Sprittles2014InertialCoalesence, Winkels2012DropletSpreading}.

The growth of the radius of the liquid bridge is compared between the immiscible and miscible droplet pairs, as shown in Figure \ref{fig:3}a. It can be found that the bridge radius of the immiscible droplets, though with a larger initial radius, grows slower than that of the miscible droplets. This difference in the bridge expansion speed shows the same trend even when the viscosity ratio between the two droplets is doubled. To further quantitatively compare the coalescence speed for miscible and immiscible droplets, the bridge expansion time that the liquid bridge grows to $0.5 R_\text{effect}$ is plotted in Figure \ref{fig:3}b. From the comparison, we can see that the bridge expansion speed of the miscible droplets is much faster than that of the immiscible droplets for all viscosity ratios. As the viscosity ratio increases, the bridge expansion time increases, indicating the coalescence of the droplets is decelerated. This deceleration, as expected, is due to the damping effect of the fluid viscosity. In addition, the difference in the bridge expansion speed between immiscible and miscible droplets is more significant when the viscosity ratio is large.

The reduction in the bridge expansion speed for immiscible droplets could be attributed to the interfacial tension of the water-oil interface. As the liquid bridge expands, the water-oil interfacial tension ${{\sigma }_{\text{wo}}}$ is pulling radially inwards at the water-oil-air three-phase contact line and resists the outwards expansion of the liquid bridge. In contrast, such force does not exist for miscible droplet pairs. Therefore, the bridge expansion speed is slower for immiscible droplet pairs than for miscible droplet pairs. The detailed force analysis will be discussed in Section \ref{sec:33}.

To further analyze the coalescence dynamics of the immiscible droplets, we consider the regimes of the coalescence process and the relevant forces. The coalescence process is initially dominated by the capillary force and the viscous force due to the low velocity. As capillary force accelerates the fluid in the liquid bridge, the fluid inertia increases while the relative importance of the viscous force decreases. Inspired by the previous studies of the coalescence dynamics of miscible droplets \cite{Chireux2021InertialCoalescence, Eggers1999DropletCoalescence, Paulson2011DropletCoalescence, Sprittles2014InertialCoalesence, Thoroddsen2005CoalescenceSpeed}, we next consider the two regimes of the coalescence of immiscible droplets, i.e., the viscous-dominated regime and the inertia-dominated regime in Sections \ref{sec:32} and \ref{sec:33}, as well as the transition between them in Section \ref{sec:34}.

\subsection{Viscous-dominated regime}\label{sec:32}
For the coalescence of miscible droplets, after the two droplets touch each other, the growth of the liquid bridge is first dominated by viscous and capillary forces \cite{Paulson2011DropletCoalescence}. Since the surrounding fluid is approximately inviscid, the viscous force results from the tangential flow at the contacting surface. The axial velocity gradient of the tangential flow is over the small longitudinal length scale, which is the bridge height $w$ and was given by Duchemin et al.\ \cite{Duchemin2003InviscidCoalescence} as $w={{{{{r}^{2}}}/{R}}_{0}}$. Then the viscous stress in the bridge can be estimated by $\mu {{{u}_{r}}}/{w}=\mu {{{u}_{r}}}/{( {{{r}^{2}}}/{{{R}_{0}}} )}$, where $u_r$ is the bridge expansion velocity ${{u}_{r}}={dr}/{dt}$. The Laplace pressure in the bridge region serves as the driving force for the bridge expansion, and can be estimated as ${\Delta P\sim \sigma {{R}_{0}}}/{{{r}^{2}}}$. From the balance of the viscous stress with the Laplace pressure, we can get a differential equation of the bridge radius
\begin{equation}\label{eq:1}
  \frac{\mu }{{{{r}^{2}}}/{{{R}_{0}}}\;}\frac{dr}{dt}\sim \frac{\sigma {{R}_{0}}}{{{r}^{2}}},
\end{equation}
which can then be integrated to give a linear growth of the liquid bridge radius:
\begin{equation}\label{eq:2}
  \frac{r}{{{R}_{0}}}=C\left( \frac{t}{{{t}_{\nu }}} \right),
\end{equation}
where $C$ is a dimensionless prefactor and ${{{t}_{\nu }} = \mu {{R}_{0}}}/{\mathsf{\sigma }}$. This linear scaling has been proved by experiments \cite{Eggers1999DropletCoalescence} and simulations \cite{Xia2019dropletCoalescence} for miscible droplets.

For the coalescence of immiscible droplets in our experiments, the Ohnesorge number (i.e., the ratio of viscous force to inertia and capillary forces, $\Oh \equiv {{{\mu }_{o}}}/{\sqrt{{{\rho }_{\text{avg}}}{{S}_{\sigma }}r}}$) at the initial regime is larger than 1 due to the small bridge radius $r$, indicating that the viscous force is dominant in the initial coalescence regime. Hence, we test whether the same linear scaling $r^*\equiv {r}/{{{R}_{0}}}\sim t^*\equiv {t}/{{{t}_{\nu }}}$ in Eq.\ (\ref{eq:2}) applies to the initial coalescence of immiscible droplets.

\begin{figure}
  \centering
  \includegraphics[scale=0.7]{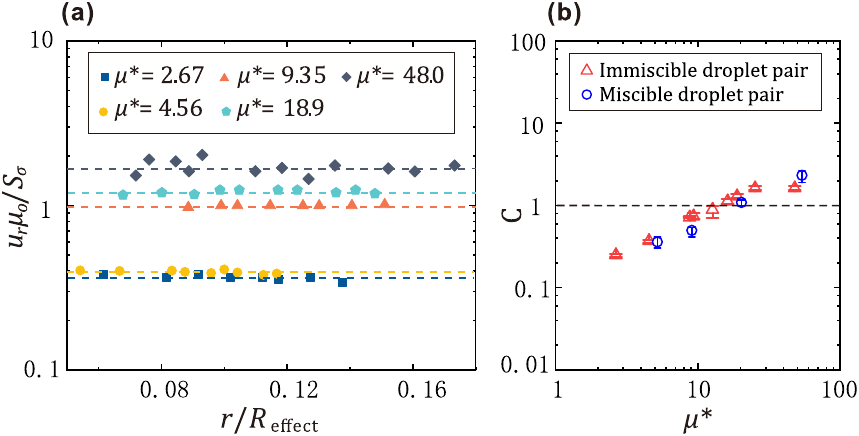}
  \caption{(a) Dimensionless velocity of the bridge expansion ${{{u}_{r}}{{\mu }_{o}}}/{{{S}_{\sigma }}}$ in the viscous-dominated regime against the dimensionless bridge radius $r/{{R}_{\text{effect}}}$. Different symbols indicate the results of different viscosity ratios. (b) Prefactors in the viscous-dominated regime \hl{of immiscible and miscible droplets} for different viscosity ratios. \hl{The error bars are the deviations from the average value of five repeated tests using the same liquids.}}\label{fig:4}
\end{figure}

To consider the initial evolution of the liquid bridge radius, we first calculate the bridge expansion velocity, ${{u}_{r}}$. We use the velocity instead of the bridge radius because the exact point of the coalescence time $t = 0$ is challenging to pinpoint at a very high resolution \cite{Chireux2021InertialCoalescence}, which will affect the plot of $r$ with $t$. The bridge expansion velocity is then nondimensionalized with the viscous capillary velocity, which for immiscible droplets can be calculated as
\begin{equation}\label{eq:3}
  {{u}_{\nu }}=\frac{{{S}_{\sigma }}}{{{\mu }_{o}}}.
\end{equation}
Figure \ref{fig:4}a shows a constant bridge expansion velocity in the initial coalescence, indicating the linear scaling $r^* \sim t^*$. Furthermore, the prefactors $C$ \hl{for miscible droplets and immiscible droplets} under different fluid viscosity ratios are shown in Figure \ref{fig:4}b, \hl{which are all in the order of $O(1)$, as was observed for coalescence of droplets of same fluids \cite{Aarts2005DropletcoalescenceHydrodynamics, Paulson2011DropletCoalescence}.}

\subsection{Inertia-dominated regime}\label{sec:33}
As the liquid bridge expands, more liquid is driven into the bridge region, and the bridge expansion velocity increases over time. Therefore, the inertia of the bridge increases, and becomes significant to dominate the flow \cite{Deka2019dropletcoalescence, Sprittles2014InertialCoalesence}. Meanwhile, the importance of the viscous force reduces. Due to the additional water-oil interfacial tension for immiscible droplet pairs, the dimensionless bridge expansion velocity (i.e., the slope in the $r^*-t^*$ plot in Figure \ref{fig:3}a) is slower than that for miscible droplets. We then study the bridge evolution of immiscible droplet pairs in the inertia-dominated regime, and analyze the effect of the droplets’ immiscibility on the coalescence process.

In the inertia-dominated regime, the expansion of the liquid bridge is dominated by the balance between inertia and capillary force. The inertia of the liquid bridge can be estimated from the second law of motion,
\begin{equation}\label{eq:4}
  {{F}_{i}}=\frac{d}{dt}\left( {{m}_{b}}{{u}_{r}} \right),
\end{equation}
where ${{u}_{r}}$ is the bridge expansion velocity, and ${{m}_{b}}$ is the mass of the liquid entrained in the liquid bridge. The mass of the bridge can be calculated from the width $w$ and the radius $r$ of the liquid bridge by assuming a cylindrical shape ${{m}_{b}}={{\rho }_{\text{avg}}}\pi {{r}^{2}}w$ (as illustrated in Figure \ref{fig:1}c), where the width of the liquid bridge $w$ can be estimated from the geometrical relation $w\sim{{{r}^{2}}}/{{{R}_{\text{effect}}}}$  \cite{Duchemin2003InviscidCoalescence}. Hence, the inertia in Eq.\ (\ref{eq:4}) can be obtained
\begin{equation}\label{eq:5}
  {{F}_{i}}\sim \frac{d}{dt}\left( \frac{\pi {{\rho }_{\text{avg}}}{{r}^{4}}}{{{R}_{\text{effect}}}}\frac{dr}{dt} \right).
\end{equation}
During the coalescence of immiscible droplets, the existence of the three-phase contact line will contribute to the capillary force. Here, in analogy to the droplet spreading process on a partial wetting surface \cite{Gennes2004Capillaritywetting}, the capillary force of the liquid bridge can be expressed as:
\begin{equation}\label{eq:6}
  {{F}_{c}}\sim 2\pi r{{S}_{\sigma }}(\cos {{\theta }_{wo}}-\cos \theta (t)),
\end{equation}
where $\theta_{wo}$ and $\theta (t)$ are the equilibrium and transient apparent contact angle of water droplet on the oil surface, which is similar to the equilibrium contact angle of a droplet on a partial wetting surface. In the present case, the three spreading parameters of the immiscible water-oil-air system are ${{S}_{w}}={{\sigma }_{oa}}-\left( {{\sigma }_{wo}}+{{\sigma }_{wa}} \right)<0$, ${{S}_{a}}={{\sigma }_{wo}}-\left( {{\sigma }_{wa}}+{{\sigma }_{oa}} \right)<0$, and ${{S}_{o}}={{\sigma }_{wa}}-\left( {{\sigma }_{wo}}+{{\sigma }_{oa}} \right)>0$, respectively. These values of the spreading parameters would lead to a total engulfment of the water phase into the oil phase if reaching the equilibrium configuration \cite{Torza1969ImmiscibleDroplets}, and also lead to an unclosed Neumann triangle when calculating the three contact angles formed among the three immiscible interfaces \cite{Gennes2004Capillaritywetting}. Therefore, the contact angle would correspond to the extreme condition, $\theta_{wo} = 0^\circ$.

The dynamic apparent contact angle $\theta \left( t \right)$ in Eq.\ (\ref{eq:6}) can be obtained from the geometry relationship \cite{Chen2013DropletSpreadingSoftSurface}, $\cos \theta (t)\sim {{{r}^{2}}}/{\left( 2R_{\text{effect}}^{2} \right)} -1$. By substituting $\cos \theta (t)$ into Eq.\ (\ref{eq:6}), we can obtain the capillary force
\begin{equation}\label{eq:7}
  {{F}_{c}}\sim 2\pi r{{S}_{\sigma }}\left( 1+\cos {{\theta }_{wo}}-\frac{{{r}^{2}}}{2{{R}_{\text{effect}}^{2}}} \right),
\end{equation}
which is the driving force for the growth of the liquid bridge. Considering the balance between the inertia $F_i$ (in Eq.\ (\ref{eq:5})) and the capillary force $F_c$ (in Eq.\ (\ref{eq:7})), we can thus obtain a governing equation of the bridge radius $r$
\begin{equation}\label{eq:8}
  k2\pi r{{S}_{\sigma }}\left( 1+\cos {{\theta }_{wo}}-\frac{{{r}^{2}}}{2{{R}_{\text{effect}}^{2}}} \right)-\frac{d}{dt}\left( \frac{\pi {{\rho }_{\text{avg}}}{{r}^{4}}}{{{R}_{\text{effect}}}}\frac{dr}{dt} \right)=0, \end{equation}
where $k$ is a scaling constant. By introducing dimensionless parameters ${{r}^{*}}\equiv {r}/{{{R}_{\text{effect}}}}$, ${{t}^{*}}\equiv {t}/{{{t}_{_{\sigma }}}}\;={t}/{\sqrt{{\left( {{\rho }_{\text{avg}}}R_{_{\text{effect}}}^{3} \right)}/{{{S}_{\sigma }}}}}$, and $\xi \equiv \sqrt{k\left[ {(1+\cos {{\theta }_{wo}})}/{\left( 2{{r}^{*}}^{2} \right)}-{1}/{4} \right]}$, Eq.\ (\ref{eq:8}) could be simplified into
\begin{equation}\label{eq:9}
  {{r}^{*}}\frac{d}{d{{t}^{*}}}\left( \frac{d{{r}^{*}}}{d{{t}^{*}}} \right)+4{{\left( \frac{d{{r}^{*}}}{d{{t}^{*}}} \right)}^{2}}-4{{\xi }^{2}}=0,
\end{equation}
which is a second-order nonlinear differential equation. After further manipulation and integration of Eq.\ (\ref{eq:9}), we can obtain
\begin{equation}\label{eq:10}
  \frac{d{{r}^{*}}}{d{{t}^{*}}}=\xi \cdot \frac{1-\exp \left[ -2\xi \left( \frac{4{{t}^{*}}}{{{r}^{*}}}+{{C}_{1}} \right) \right]}{1+\exp \left[ -2\xi \left( \frac{4{{t}^{*}}}{{{r}^{*}}}+{{C}_{1}} \right) \right]},
\end{equation}
where $C_1$ is a constant produced during the integration.

\begin{figure}
  \centering
  \includegraphics[scale=0.9]{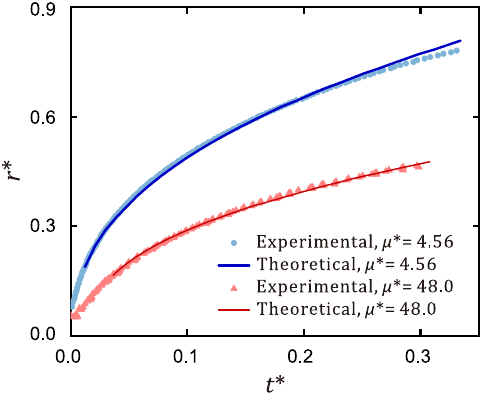}
  \caption{Comparison of experimental results and theoretical results. The bridge radius and the time are nondimensionalized as $r^* \equiv r/R_\text{effect}$ and $t^* \equiv t/t_\sigma$, respectively.}\label{fig:5}
\end{figure}

Eq.\ (\ref{eq:10}) can be solved by numerical integration while taking the bridge radius in the inertia-dominated regime as the initial condition. We used MATLAB ode45 to integrate Eq.\ (\ref{eq:10}) for two typical cases and compared the results with the experimental data (see Figure \ref{fig:5}). The comparison
shows that the theoretical model is in good agreement with the experimental data.

It should be noted current model (Eq.\ (\ref{eq:10})) can also be used for miscible droplet pairs.
As there is no immiscible interface between the two miscible droplets during their coalescence process, the capillary force that drives the motion of the liquid bridge is
  ${{F}_{c}}\approx 2\pi r\sigma$,
where $\sigma$ is the surface tension of the droplet or the lower value if the surface tension of the two droplets is different.
Hence, the dimensionless parameter $\xi$ in Eqs.\ (\ref{eq:9}) and (\ref{eq:10}) is reduced to $\xi = \sqrt{{k}/{\left( 2{{r}^{*}}^{2} \right)}}$.
Previous studies have shown that the growth of the liquid bridge follows a 1/2 power law in the inertia-dominated regime \cite{Chireux2021InertialCoalescence, Paulson2011DropletCoalescence, Thoroddsen2007InitialDropletCoalescence}. The 1/2 power law was developed by assuming the balance between the surface tension stress ${\sigma {{R}_{0}}}/{{{r}^{2}}}$ and the dynamic pressure $\rho {{v}^{2}}$. By setting the velocity $v = {dr}/{dt}$, the 1/2 power law can be obtained:
\begin{equation}\label{eq:s1}
  r\propto {{\left( \frac{\sigma {{R}_{0}^{3}}}{\rho } \right)}^{{1}/{4}}}{{t}^{{1}/{2}}}.
\end{equation}

\begin{figure}
  \centering
  \includegraphics[scale=0.75]{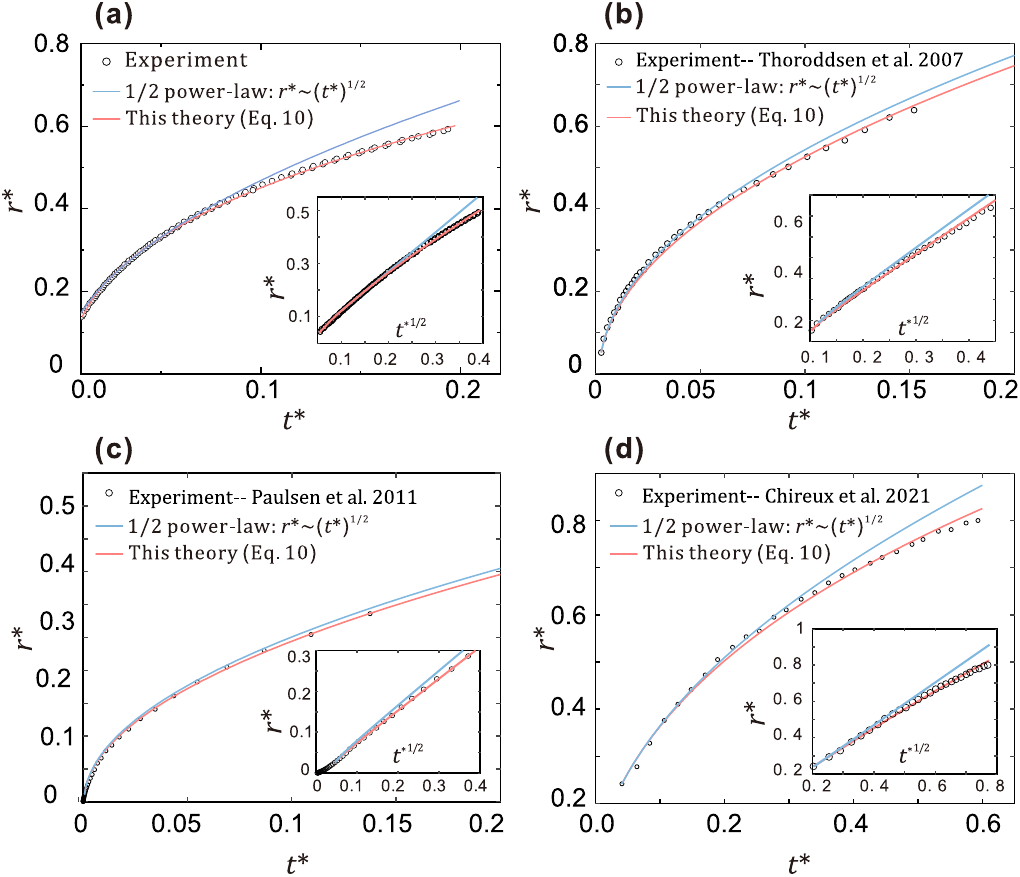}
  \caption{Comparison of the current theoretical model in Eq.\ (\ref{eq:10}) in bridge evolution of miscible droplets with experimental data: (a) our experimental data (coalescence of a water droplet with an ethanol-glycerol mixture droplet with ${{\mu }^{*}}=9.09$); (b) experimental data in Fig.\ 6 in Ref.\ \cite{Thoroddsen2007InitialDropletCoalescence} for an ethanol droplet (90\% weight concentration) with a water droplet; (c) experimental data in Fig.\ 2 in Ref.\ \cite{Paulson2011DropletCoalescence} for two droplets of glycerol-water mixture (viscosity is 1.9 mPa$\cdot$s); (d) experimental data in Fig.\ 10 in Ref.\ \cite{Chireux2021InertialCoalescence} for two water--MgSO$_4$ droplets. The radius of the liquid bridge is nondimensionalized by the initial effective droplet radius ${{r}^{*}} = r/{{R}_\text{effect}}$, and the time is nondimensionalized by the capillary-inertia time scale $t^* =  t/{{t}_{\sigma }}=t/\sqrt{{{{\rho }_{\text{avg}}}R_{\text{effect}}^{3}}/{\sigma }}$. \hl{Inset: rescale of the nondimensional time axis with the 1/2 power law.}}
  \label{fig:s2}
\end{figure}

Comparison is made between the solution of Eq.\ (\ref{eq:10}) and the 1/2 power law in Eq.\ (\ref{eq:s1}) with our experimental data (see Figure \ref{fig:s2}a) and the experimental data in the literature \cite{Chireux2021InertialCoalescence, Paulson2011DropletCoalescence, Thoroddsen2007InitialDropletCoalescence} (see Figure \ref{fig:s2}b-d). We can see that both two models agree well with the experimental data in the early inertial-dominated regime, \hl{We also plot the dimentionless bridge radius against ${t^*}^{1/2}$ as shown in the insets of Figure \ref{fig:s2}, and the deviation from a line of constant slope further indicates that the liquid bridge growth would fail to comply with the conventional 1/2 power law growth in a longer timescale.} Such deviation is consistent with the present theory of Eq.\ (\ref{eq:10}) and can be explained  by the fact that the force balance in the 1/2 power law model does not take the mass variation of the liquid bridge into account. As the liquid bridge expands, more liquid is entrained into the bridge region. By considering the mass variation of the liquid bridge based on momentum conservation, our model can better capture the evolution of the liquid bridge in the inertia-dominated regime. Therefore, the current model is not only applicable to the coalescence of miscible droplets, but also could better capture the coalescence dynamics in a longer time than the 1/2 power law.

\subsection{Transition from viscous-dominated regime to inertia-dominated regime}\label{sec:34}
We then consider the transition from the initial viscous-dominated regime to the inertia-dominated regime. The transition should occur when the inertia stress and the  viscous stress in the bridge area are comparable with each other. For the coalescence of immiscible droplets, as discussed in Section \ref{sec:32}, the dominant viscous stress is estimated as ${{\mu }_{o}}\left( {{{u}_{r}}}/{w} \right)={{\mu }_{o}}\left[ {{{u}_{r}}}/{\left( {{{r}_{c}^{2}}}/{{{R}_{\text{effect}}}} \right)} \right]$, and the dominant inertia stress in the bridge region can be estimated as ${{\rho }_{\text{avg}}}{{u}_{r}^{2}}$. To estimate the transition condition, the bridge expansion velocity ${{u}_{r}}$ can be estimated by the viscous capillary velocity, which, for immiscible droplet pairs, is ${{u}_{r}}={{{S}_{\sigma }}}/{{{\mu }_{o}}}$. Therefore, by comparing the viscous stress and the inertia stress, we can obtain a modified Ohnesorge number for the transition
\begin{equation}\label{eq:11}
  {{\Oh}_{T}}=\sqrt{\frac{{{\mu }_{o}^{2}}{{R}_{\text{effect}}}}{{{S}_{\sigma }}{{\rho }_{\text{avg}}}{{r}_{c}^{2}}}}.
\end{equation}

\begin{figure}
  \centering
  \includegraphics[scale=0.75]{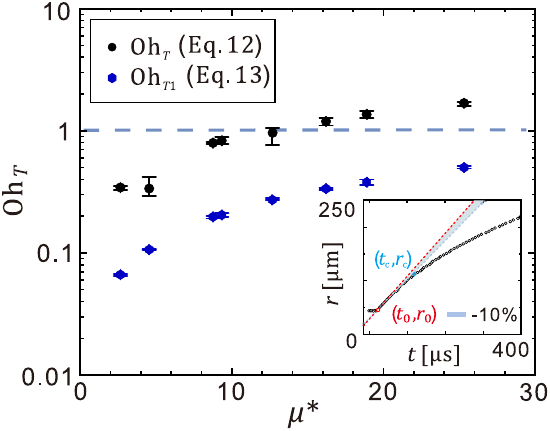}
  \caption{Transition Ohnesorge number for different viscosity ratios. The black points show the Ohnesorge number $\Oh_{T}$ in Eq.\ (\ref{eq:11}) based on our analysis, while the blue points show the transition Ohnesorge number $\Oh_{T1}$ in Eq.\ (\ref{eq:12})  based on the analysis by Aarts et al.\ \cite{Aarts2005DropletcoalescenceHydrodynamics}. The inset shows the \hl{identification of time zero and} the determination of the transition radius. \hl{The error bars are the deviations from the average value of five repeated tests using the same liquids.}}\label{fig:6}
\end{figure}

To find the transition radius ${{r}_{c}}$ in Eq.\ (\ref{eq:11}) from the experimental data, \hl{we first obtained the line of linear growth ${ r}={r}_{0}+{A}({t}-{t}_{0})$ by fitting using the method adopted by Dekker et al. \cite{Dekker2022ElacticityDroplet},}
and \hl{then} identified ${{r}_{\text{c}}}$ as the first point departing more than --10\% from this line of constant velocity, as sketched in the inset of Figure \ref{fig:6}. By plotting the transition Ohnesorge number $\Oh_T$ using the obtained transition radius, we can find all the values of $\Oh_T$ are in the order of unity as expected, as is shown in Figure \ref{fig:6}. For comparison, another transition Ohnesorge number $\Oh_{T1}$ based on the analysis of Aarts et al.\ \cite{Aarts2005DropletcoalescenceHydrodynamics} was also plotted in Figure \ref{fig:6},
\begin{equation}\label{eq:12}
  {{ \Oh}_{T1}}=\frac{{{\mu }_{o}}}{\sqrt{{{S}_{\sigma }}{{r}_{c}}{{\rho }_{\text{avg}}}}}.
\end{equation}
$\Oh_{T1}$ was obtained by considering the bridge radius as the dominant viscous length scale and estimating the viscous stress as ${{{\mu }_{o}}{{u}_{r}}}/{r}$. As shown in Figure \ref{fig:6}, $\Oh_{T1}$ for different viscosity ratios are much less than $O(1)$. This result confirms that $\Oh_{T}$ in Eq.\ (\ref{eq:11}) can be used to characterize the transition from the viscous-dominated regime to the inertia-dominated regime in the coalescence of immiscible droplets.

\section{Conclusion}\label{sec:4}
In conclusion, unlike previous studies on the liquid bridge dynamics during droplet coalescence, which only considered miscible droplets \cite{Aarts2005DropletcoalescenceHydrodynamics, Burton2007Dropletpinchoff, Eggers1999DropletCoalescence, Thoroddsen2005CoalescenceSpeed, Wu2004scalinglawCoalescence, Xia2019dropletCoalescence}, this work, for the first time, investigates the growth of the liquid bridge of immiscible droplets and shows a detailed insight into the effect of immiscibility on the bridge evolution. In the early viscous-dominated regime, the bridge dynamics of immiscible droplets is found to show a similar linear growth as that of the miscible droplets \cite{Rahman2019ViscousCoalescence, Xia2019dropletCoalescence}. While in the later inertia-dominated regime, the bridge dynamics is observed to be affected by the immiscible interfacial tension and exhibits different growth dynamics. In this regime, the bridge growth is slowed by the pulling effect of the water-oil interfacial tension \cite{Bernard2020DropletImpactFilm, Thoraval2013AirContraction}. \hl{Based on the balance between inertia and the capillary force considering the existence of the three-phase contact line}, a theoretical model is proposed to predict the growth of the liquid bridge, which agrees well with our experimental data of immiscible droplets. In addition, this model could also be used for the prediction of the bridge growth of miscible droplets, and the results demonstrate that it could better predict the bridge evolution in a longer period compared to the conventional 1/2 power law in the inertia-dominated regime \cite{Eggers1999DropletCoalescence, Thoroddsen2005CoalescenceSpeed, Wu2004scalinglawCoalescence, Xia2019dropletCoalescence}, \hl{as the developed model takes the mass variation of the liquid bridge into consideration.} Finally, the comparison of dominant \hl{stress} in the two regimes gives a modified Ohnesorge number, which can be used to determine the transition from the viscous-dominated regime to the inertia-dominated regime.

The bridge expansion dynamics of immiscible droplets established in this study not only serves as a complement to previous work on miscible droplet coalescence, but also helps gain a deeper insight into droplets coalescence process involving immiscible liquids and facilitates related applications, with examples such as the accurate manipulation of droplets in material and biological applications \cite{Shen2015MicroDropletCoalescence}. This study considers the coalescence of immiscible droplets without any external force (i.e., the gravitational force is negligible). External forces such as electric fields \cite{2022HartmannCoalescenceElectricField, Sadeghi2018ChargedDropCoalescence} or magnetic fields \cite{Edward2019MosesEffect, He2020MagneticNanoparticles} can be used to actively control the coalescence process to facilitate the relevant applications, and the effects of these forces on the coalescence dynamics could be studied in the future.

\section*{Declaration of Competing Interest}
The authors declare that they have no known competing financial interests or personal relationships that could have appeared to influence the work reported in this paper.

\section*{Acknowledgements}
This work is supported by the National Natural Science Foundation of China (Grant Nos.\ 52176083 and 51921004).

\section*{Appendix A. Supplementary material}
Supplementary material associated with this article can be found in the online version.

\bibliography{ImmiscibleCoalescence}

\end{document}